\documentclass[acus]{JAC2000}
\usepackage{graphics}

\providecommand{\LyX}{L\kern-.1667em\lower.25em\hbox{Y}\kern-.125emX\@}

\makeatother
\begin{document}

\setlength{\titleblockheight}{60mm}

\title{\flushright{WEAP051}\\[15pt] \centering Development of the Simulator for the Global Orbit Feedback System using
EPICS}

\author{K. H. Kim\thanks{
kimkh@postech.ac.kr, http://fal.postech.ac.kr
}, J. Choi, T-Y Lee, J. H. Kim, Yujong Kim, C. Kim, Guinyun Kim\( ^{1} \), \\M.
H. Cho, W. Namkung, and I. S. Ko\thanks{
isko@postech.ac.kr
} \\\\Pohang Accelerator Laboratory, POSTECH, Pohang 790-784, Korea
\\\( ^{1} \)Center of High Energy Physics, Kyungpook National University,
Daegu 702-701, Korea}

\maketitle
\begin{abstract}
We have carried out the basic research for the accelerator and tokamak
control system based on the Experimental Physics and Industrial Control
System (EPICS). We have used the process database and the state notation
language (SNL) in the EPICS to develop the simulator which represents
as a virtual machine. In this paper. In this paper, we introduce the
simulator of the global orbit feedback system as an example. This
simulates the global orbit feedback system under the constraint conditions
for Pohang Light Source (PLS) storage ring. We describe the details
of the feedback algorithm and the realization of the simulator.
\end{abstract}

\section{Introduction}

When the beam position change takes place during the conventional
global orbit correction processes, the photon beam through the beamline
is affected, and it results in the alignment of mirrors and monochromators.
This is particularly severe for a long beamline such as the undulator
beamlines. This problem can be overcome by introducing a local bump
at the particular beamline. However, for some light sources, there
are not enough corrector magnets to generate local bumps as much as
needed. This difficulty can be overcome when we correct the COD under
the condition where the beam positions at particular points are not
changed. This is our main objective to develop a method of the closed
orbit correction under constraint conditions.

\section{Theory }

\subsection{Ordinary COD correction and regularization}

The beam position is normally described as a vector $\left| {\rm\bf x} \right>$
measured by $M$ beam position monitors (BPM). In order to correct
the COD, we need $N$ corrector magnets with their strengths described
as a vector $\left| {\rm\bf k} \right>$. When the corrector magnets
kick a beam, the new beam positions $\left| {\rm\bf y} \right>$ can
be described as follows.\begin{eqnarray} \left| {\rm\bf y} \right> = {\rm\bf R} \left| {\rm\bf k} \right> + \left| {\rm\bf x} \right> . \end{eqnarray} \noindent
Here, ${\rm\bf R}$ is called the response matrix of $(M \times N)$
dimensions whose components are given by \begin{eqnarray} {\rm\bf R}_{ij} = \frac{\sqrt{\beta_i \beta_j}}{2 \sin \pi \nu} \cos \left(\left| \Psi_i - \Psi_j \right| - \pi \nu \right) , \end{eqnarray} \noindent
where $\nu$ is the betatron tune of the storage ring, and $(\beta_i, \Psi_i)$
and $(\beta_j, \Psi_j)$ are the beta function and the phase function
for the $i^{th}$ BPM and $j^{th}$ corrector magnet, respectively.
In order to reduce the COD, we have to choose the kick of each corrector
magnet satisfying \begin{eqnarray} {\rm\bf R}^T {\rm\bf R} \left| {\rm\bf k} \right> + {\rm\bf R}^T \left| {\rm\bf x} \right> =0~. \label{coc_eqn_ordinary_COD_relation} \end{eqnarray} \noindent
It is called the PSINOM algorithm\cite{coc_PSINOM}. 

The COD correction is actually a minimization procedure of $S$ defined
as \begin{eqnarray} S=\frac{1}{2} \left\{ \left< {\rm\bf k} \right| {\rm\bf R}^T {\rm\bf R} \left| {\rm\bf k} \right> + 2 \left< {\rm\bf x}\right| {\rm\bf R} \left| {\rm\bf k} \right> + \left< {\rm\bf x} \right. \left| {\rm\bf x} \right> \right\} . \end{eqnarray} \noindent
By using the relations of vector operators which are hyper-dimensional
gradient operators, we can get the same result of PSINOM algorithm
as shown in eq. (\ref{coc_eqn_ordinary_COD_relation}).

This algorithm includes an inversion procedure of the matrix ${\rm\bf R}^T {\rm\bf R}$.
In some cases, we can get unacceptable corrections due to the ill-posedness
of ${\rm\bf R}^T {\rm\bf R}$. A regularization method is introduced
to avoid this problem. In this case, $S$ is written as \begin{eqnarray} S=\frac{1}{2} \left\{ \left< {\rm\bf k} \right| {\rm\bf R}^T {\rm\bf R} \left| {\rm\bf k} \right> + 2 \left< {\rm\bf x}\right| {\rm\bf R} \left| {\rm\bf k} \right> + \left< {\rm\bf x} \right. \left| {\rm\bf x} \right> \right\} \cr + \frac{1}{2} \left< {\rm\bf k} \right| \alpha \left| {\rm\bf k} \right> . \end{eqnarray} \noindent
where $\alpha$ is the regularization parameter. Then the minimum corrector
kicks can be determined by \begin{eqnarray} \left( {\rm\bf R}^T {\rm\bf R} + \alpha {\rm\bf I} \right) \left| {\rm\bf k} \right> + {\rm\bf R}^T \left| {\rm\bf x} \right> =0~. \label{coc_eqn_modified_PSINOM} \end{eqnarray} \noindent
This equation represents the modified PSINOM algorithm, and we can
relax the inversion problem of the singular matrix by using the diagonal
matrix $\alpha {\rm\bf I}$ when ${\rm\bf R}^T {\rm\bf R}$ is singular\cite{coc_regularization}.

\subsection{Method with constraint conditions}

After having the new closed orbit with the minimum distortion from
eq. (\ref{coc_eqn_modified_PSINOM}), the new orbit is generally different
from the original orbit. Sometimes, this difference can be taken place
at very sensitive locations such as the entrance and the exit of an
undulator. If the beamline is well aligned for this undulator, a COD
correction should be avoided in this region. A constraint condition
can be described in terms of the beam position at $i^{\rm th}$ BPM
such as \begin{eqnarray} \left< {\rm\bf R}_i \right. \left| {\rm\bf k} \right> + x_{0i} = x_{i} ~. \end{eqnarray}

\noindent Here, $\left< {\rm\bf R}_i \right|$ is the $i^{\rm th}$
row of the response matrix ${\rm\bf R}$. Also, $x_{0i}$ and $x_i$
are the beam positions before and after the correction, respectively.
Since we want to keep this position unchanged, $\left< {\rm\bf R}_i \right. \left| {\rm\bf k} \right>$
should be zero. If there are $L$ BPMs involved in the constraint condition,
we can write the constraint condition as follows. \begin{eqnarray} {\rm\bf C}^T \left| {\rm\bf k} \right> =0 . \label{coc_eqn_constraint} \end{eqnarray} \noindent
Here, ${\rm\bf C}^T$ is the ($L \times N$) sub-matrix of the response
matrix. Each component of ${\rm\bf C}^T$ corresponds to the BPM involved
in the constraint condition. We also assume that $\left| {\rm\bf k} \right>$
has a non-trivial solution.

We now add this constraint condition to the modified PSINOM algorithm
to obtain the new $S$ such as, \begin{eqnarray} S=\frac{1}{2} \left[ \left< {\rm\bf k} \right| {\rm\bf R}^T {\rm\bf R} \left| {\rm\bf k} \right> +2 \left< {\rm\bf x} \right| {\rm\bf R} \left| {\rm\bf k} \right> + \left< {\rm\bf x} | {\rm\bf x} \right> \right] \cr +\frac{1}{2} \left< {\rm\bf k} \right| \alpha \left| {\rm\bf k} \right> +\left< {\rm\bf \Gamma} \right| {\rm\bf C}^T \left| {\rm\bf k} \right> . \end{eqnarray} \noindent
Here, $\left< {\rm\bf \Gamma} \right|$ is the Lagrangian multiplier,
and it is an $L$ dimensional vector. By following the derivative to
the corrector strength, we can get the vector $\left| {\rm\bf k} \right>$
which minimizes the closed orbit distortion outside the constraint
region such as, \begin{eqnarray} \left( {\rm\bf A} + \alpha {\rm\bf I} \right) \left| {\rm\bf k} \right> + {\rm\bf R}^T \left| {\rm\bf x} \right> + {\rm\bf C} \left| {\rm\bf\Gamma} \right> =0 . \label{coc_eqn_new_correction} \end{eqnarray} \noindent
Here, we define the square matrices of $N \times N$ dimensional ${\rm\bf A}$
and $L \times L$ dimensional ${\rm\bf D}$ as follows. \begin{eqnarray} {\rm\bf A} = {\rm\bf R}^T {\rm\bf R} , \\ {\rm\bf D}= {\rm\bf C}^T \left( {\rm\bf A} + \alpha {\rm\bf I} \right)^{-1} {\rm\bf C} ~. \end{eqnarray} \noindent
Eq. (\ref{coc_eqn_new_correction}) can be rewritten as \begin{eqnarray} \left| {\rm\bf\Gamma}\right> = -{\rm\bf D}^{-1} {\rm\bf C}^T \left( {\rm\bf A} + \alpha {\rm\bf I} \right)^{-1} \left| {\rm\bf x} \right> ~. \end{eqnarray} \noindent
Now, we can remove the Lagrangian multiplier $\left| {\rm\bf \Gamma} \right>$
in eq. (\ref{coc_eqn_new_correction}) by using the above equation.
Then, we can finally get the kick values of the corrector magnets
as follows. \begin{eqnarray} \left| {\rm\bf k} \right> = - \left( {\rm\bf A} + \alpha {\rm\bf I} \right)^{-1} \left\{ {\rm\bf I} - {\rm\bf C} {\rm\bf D}^{-1} {\rm\bf C}^T \left( {\rm\bf A} + \alpha {\rm\bf I} \right)^{-1} \right\} \cr {\rm\bf R}^T \left| {\rm\bf x} \right> ~. \label{coc_eqn_final_eq} \end{eqnarray}

\section{Development of simulator using EPICS}

The algorithm developed in the previous section has been successfully
tested for the Pohang Light Source (PLS) operation. The new correction
code is written in C language and it is installed in one of the operator
consoles which is a SUN workstation. Although the PLS control system
is not using Experimental Physics and Industrial Control System (EPICS)
which is used in many accelerator laboratories world-widely, there
is a plan to upgrade the control system based on EPICS technology\cite{EPICSHOME, IOCAPP}.
As a part of such upgrade activity, we have started the development
orbit correction algorithm with EPICS. Since we do not have any EPICS-based
control system yet, we decide to develop the orbit correction simulator
using EPICS.

In order to seek the way to adopt our orbit correction algorithm into
EPICS, we have considered two ways: one is based on the subroutine
record and the other is using the state notation language (SNL) program.
The first method needs new record support that runs the orbit correction
algorithm. To do this, the correction code is required to be written
upon the protocol required by the record support such as the entry
structure and the callback structure. This approach gives relatively
fast response because this method uses database access and can access
the record by process passive mode\cite{IOCAPP, RECORD}. This is
a good feature for the realtime system but the large portion of the
orbit correction code we have already tested must be rewritten according
to the protocol of the record support.

One of important tasks of EPICS input/output controller (IOC) is the
sequencer that runs programs written in SNL. The SNL considers the
control object as the state machine and treats transitions between
states. The sequencer monitors the transitions for the SNL and runs
callback functions using the entry table the corresponding program
written in SNL. Since the sequencer accesses to the record via channel
access (CA) and the record access is only possible through non-process
passive mode, there are some restrictions in access time or the treatment
of records. However, this method can directly imbed the program written
in C language. Also, unlike the subroutine record, this method can
remove program tasks without rebooting the system, which gives the
code debugging very easy\cite{IOCAPP, SNL}. Thus, the second method
gives more benefits when the system does not require heavy realtime
demands. Upon reviewing the two methods, we have decided to use the
latter method.

\begin{figure*}
{\centering \resizebox*{0.52\textwidth}{!}{\includegraphics{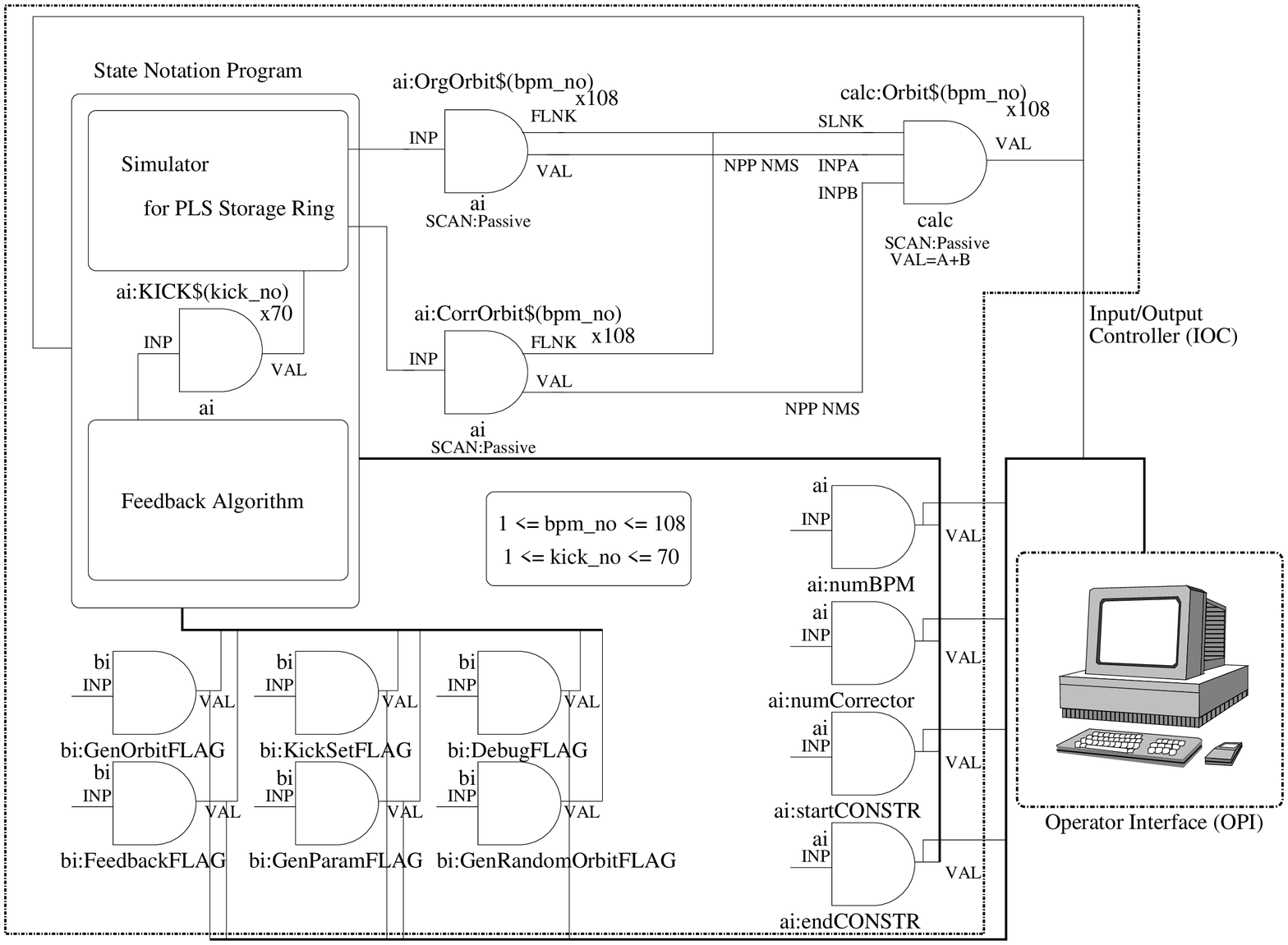}} \par}

\caption{Schemematic diagram of the orbit correction simulator\label{fig_schemetic}}
\end{figure*}

\subsection{SNL program}

The orbit correction simulator is developed in IOC level and has the
SNL program imbedding C codes and several database records, as shown
in Fig. \ref{fig_schemetic}. There are two parts in the SNL program:
one for the feedback including the orbit correction algorithm and
the other for the simulator which emulates the PLS storage ring. The
latter calculates orbit changes from {\it ai:KICK\$(kick\_no)} records
which store the corrector strengths obtained from the response matrix
measured at the PLS. This process is actually a product of matrix
and column vectors. The calculated orbit changes are stored at {\it ai:CorrOrbit\$(bpm\_no)}
records. In the next step, the corrector strengths s calculated by
the algorithm described in the previous section and the results are
stored at {\it al:KICK\$(kick\_no)}.

\subsection{Database and record linkage}

The SNL program we have developed consists of four state sets. Since
they are linked together and interacted dynamically via EPICS database,
we need to understand the database and its records as well as the
linkage between them. The {\it ai:KICK\$(kick\_no)} record uses the
analog input record to represent the corrector strength. There are
70 records altogether as representing 70 correctors in the PLS storage
ring. They are linking the simulator and the feedback in a parallel
manner. The index {\it \$(kick\_no)} is the integer value between
one and 70.

The records of {\it ai:OrgOrbit\$(bpm\_no)} and {\it ai:CorrOrbit\$(bpm\_no)}
are the orbit distortion before the feedback is applied and the orbit
change after the feedback, respectively. They are using analog input
records. Since these records represents the changes at the PLS BPM,
{\it \$(bpm\_no)} is the interger value between one and 108. The superposition
of these two records gives new orbit. This can be done by the calculation
record {\it calc:Orbit\$(bpm\_no)}. Since this calculation record
is linked forwardly from {\it ai:OrgOrbit\$(bpm\_no)} and {\it ai:CorrOrbit\$(bpm\_no)},
the superposition is newly calculated whenever the values of two records
are changed.

On the other hand, {\it ai:numBPM}, {\it ai:numCorrector}, {\it ai:startCONSTR},
and {\it ai:endCONSTR} are representing the BPMs and correctors used
in the correction algorithm, start and end point of the constraint
region, respectively. They are using analog input records and notifying
the SNL program if necessary.

The other records are binary input records and they are representing
necessary status flags. They are used as the mediator of state transitions
between state sets in the SNL program.

\section{Conclusion}

We have developed the COD correction algorithm under the constraint
condition where the beam position at particular point is not changed.
The new algorithm is based on the modified PSINOM algorithm which
includes the regularization process in order to avoid the inversion
problem of the ill-posed response matrices.

We have confirmed that this algorithm is working well and is in good
agreement with the experimental results \cite{COCPAPER}. Even though
the PLS is planning to upgrade their control system with EPICS, there
is no working EPICS based control system at PLS. Due to this, we have
developed the orbit correction simulator using C-code embedded SNL
program based on EPICS technology. This simulator part can be replaced
by the real control system with minor changes after the completion
of the upgrade.


\begin{thebibliography}{1}
\bibitem{coc_PSINOM}W. Herr: \char`\"{}Algorithms and procedures used in the orbit correction
package COCU,\char`\"{} CERN SL/95-07 (AP), 1995. 
\bibitem{coc_regularization}Y. N. Tang and S. Krinsky: Proc. AIP Conf. {\bf 315} (AIP Press, 1993)
87. 
\bibitem{EPICSHOME}See URL\char`\"{}{\it http://www.aps.anl.gov/epics}\char`\"{}
\bibitem{IOCAPP}Martin R. Kraimer: \char`\"{}EPICS IOC Application Developer's Guide,\char`\"{}
APS/ANL, 1998
\bibitem{SNL}Andy Kozubal: \char`\"{}State Notation Language and Sequencer User
Guide,\char`\"{} LANL, 1995
\bibitem{RECORD}Philip Stanley, {\it et al.}: \char`\"{}EPICS Record Reference Manual,\char`\"{}
LANL, APS/ANL, 1995
\bibitem{COCPAPER}Kukhee Kim, Jinhyuk Choi, Tae-Yeon Lee, Guinyun Kim, Moohyun Cho,
Won Namkung, In Soo Ko: Jpn. J. Appl. Phys. {\bf 40} (2001) 4233.\end{thebibliography}
\end{document}